\documentclass[11pt,twoside]{article}
\usepackage{asp2010}
\usepackage{epsf}
\usepackage{psfig}

\resetcounters

\begin{document}

\title{Hybrid pulsators among A/F-type stars}
\author{G. Handler,$^{1,2}$ 
\affil{$^1$Copernicus Astronomical Center, Bartycka 18, 00-716 Warsaw, 
Poland}
\affil{$^2$Institut f\"ur Astronomie, Universit\"at Wien, 
T\"urkenschanzstra\ss e 17, 1180 Wien, Austria}}

\begin{abstract}
This article reviews the present knowledge on oscillating main sequence 
A/F stars that belong to more than one class of pulsator. Due to recent 
results of asteroseismic space missions, we now know of many 
$\delta$~Scuti/$\gamma$~Doradus stars. However, $\gamma$~Doradus 
variability was also detected in a rapidly oscillating Ap star, and 
solar-like oscillations were discovered in a $\delta$~Scuti star. The 
astrophysical information that can be gained from these pulsators is 
discussed, and confronted with what is believed to be known about 
pulsational driving.
\end{abstract}

\section{Introduction}

Over the last 40 years, our knowledge about pulsating stars has greatly 
expanded. The ever increasing accuracy in stellar observations and
advances in theory have revealed a large number of previously unknown 
classes of variable star. Figure 1 compares the knowledge on the 
positions of pulsating stars in the HR diagram over this time span.

\begin{figure}
\plotfiddle{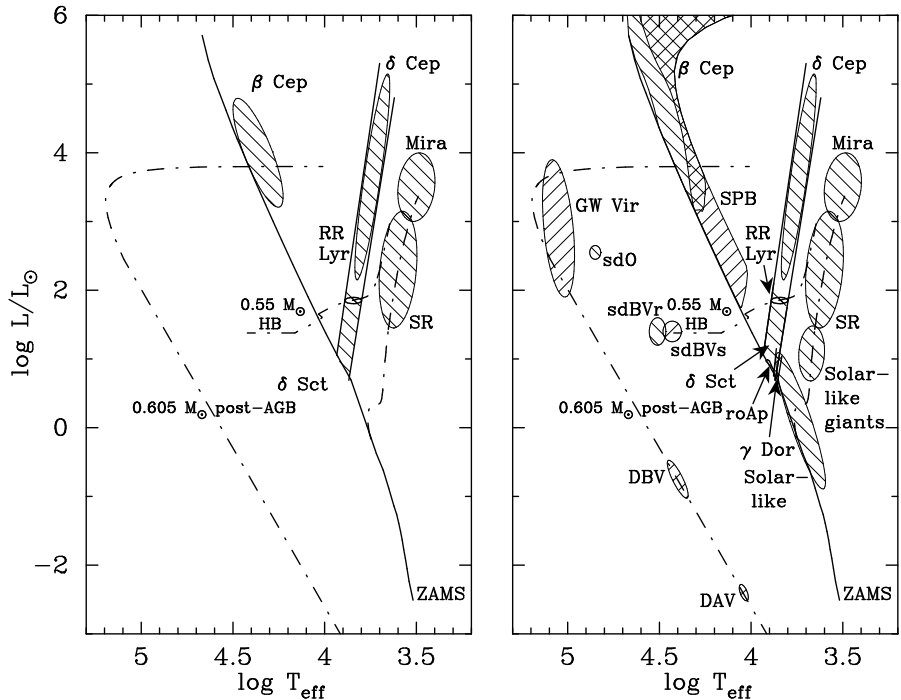}{8.3cm}{00}{72}{72}{-203}{-40}
\caption{Left: classes of pulsating star known in the late 1960's. 
Right: a selection of classes of pulsating star known to date. Areas 
hatched from lower right to upper left depict domains of g-mode 
pulsators, areas hatched from lower left to upper right delineate 
domains of p-mode pulsators; overlapping areas may contain "hybrid" 
pulsators. Parts of model evolutionary tracks for main sequence, 
horizontal branch and post-AGB stars are shown as dashed-dotted lines 
for orientation.}
\end{figure}

The most striking difference between the two panels in this figure is 
that the HR Diagram nowadays is well filled with pulsating variables. 
This has the important consequence that stellar oscillations can be 
utilized to sound the interiors of stars in many different evolutionary 
stages, and over wide ranges of mass and chemical composition. Besides 
better defined locations of the pulsating stars over the last 40 years, 
another new feature is obvious: several of the different instability 
strips overlap.

These instability strips contain stars with different pulsational 
behaviour, or in other words, with different types of mode excited. 
Consequently, it can be suspected that stars possessing more than a 
single set of mode spectra exist. Ground based observations have shown 
that they do, and the history of their study has been reviewed elsewhere 
(e.g., Handler 2009).

The most complicated region in the HR Diagram in terms of overlap of 
different classes of pulsating star is where the lower classical 
instability strip intersects the main sequence, at spectral types of A 
and F. There one finds:
\begin{itemize}
\item $\delta$ Scuti stars, pulsating in radial and nonradial pressure 
and mixed modes of low radial order with periods between 20 
minutes and 6 hours, driven by the classical $\kappa$ mechanism in the 
He II ionization zone (Baker \& Kippenhahn 1962)
\item rapidly oscillating Ap (roAp) stars, high-order 
pressure modes pulsators governed by a global magnetic field, with 
periods of 5 - 20 minutes, believed to be driven by the $\kappa$ 
mechanism in the He I/H ionization zone (Balmforth et al.\ 2001)
\item $\gamma$ Doradus stars that oscillate in high-order g modes with 
periods between 0.3 - 3 days, and likely driven by blocking of flux at 
the base of the envelope convection zone (Guzik et al.\ 2000)
\item solar-like oscillations with periods around 10 - 40 minutes are 
theoretically predicted to be present and driven by surface convection 
(Houdek et al.\ 1999)
\end{itemize}

The present article reviews the present knowledge on hybrid pulsators in 
this domain in the light of recent results of asteroseismic space 
missions. There will be a strong focus on Kepler data, not only because 
they have the best quality, but for the practical reason that the author 
is involved in their analysis.

\section{Delgam Scudor stars}

When searching for $\delta$~Scuti pulsations in $\gamma$~Doradus stars, 
Bob Shobbrook invented the term ``Delgam Scudor stars'' for such hybrids. 
The asteroseismic observations by the Kepler satellite (Gilliland et 
al.\ 2010) revealed a large number of such variables. An example light 
curve is shown in Fig.\ 2.

\begin{figure}
\plotfiddle{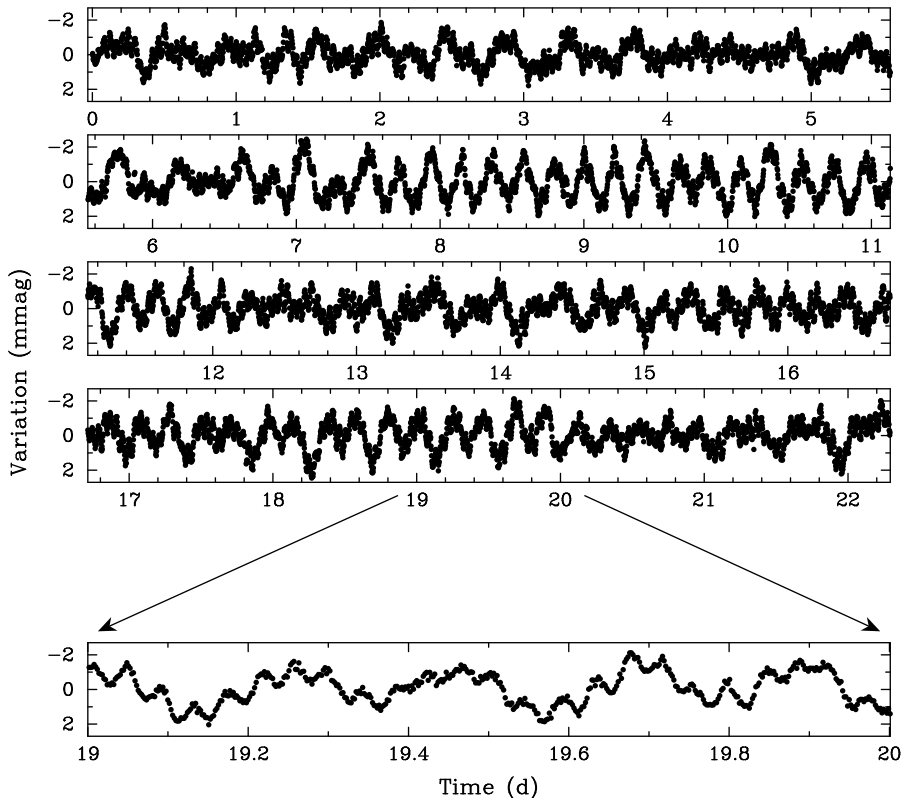}{9.5cm}{00}{70}{70}{-203}{-40}
\caption{Upper four panels: three weeks of Kepler data of a Delgam 
Scudor star. Light variations due to beating of multiperiodic g mode 
pulsations are clearly visible. Lower panel: zoom into a one-day segment of these 
measurements, highlighting the simultaneously present short-period p 
mode oscillations.}
\end{figure}

The presence of multiperiodic pulsation with both long and short periods 
is quite obvious thanks to the excellent data quality. First analyses of 
Kepler data (Grigahc\`ene et al.\ 2010, Uytterhoeven et al.\ 2011) 
revealed that $\delta$ Scuti/$\gamma$ Doradus hybridity is quite common 
(with an incidence of about 1/4) among A/F type stars, a result 
consistent with findings from CoRoT (Hareter et al.\ 2010).

Of course, better data also provide new challenges. It was noticed that 
the instability domains of $\delta$ Scuti and $\gamma$ Doradus stars, 
and consequently, such hybrids, are larger than previously thought. This 
is in general not a surprise, but the apparent presence of hybrids all 
over the $\delta$ Scuti instability strip (Grigahc\`ene et al.\ 2010, 
Uytterhoeven et al.\ 2011), and beyond, is. However, it must be kept in 
mind that the effective temperatures and surface gravities of these 
faint stars are not yet known to high accuracy. Work is in progress to 
improve this situation.

Another new result from the space missions is that the g and p mode 
domains in many Delgam Scudor stars are not nicely separated in 
frequency, as to be expected from their defining pulsational behaviour. 
Several stars appear to have a continuum of pulsation frequencies 
excited, filling the gap between high-order g modes and low order p and 
mixed modes.

This is a problem for models of pulsational mode excitation, which 
predict just such a clear separation (e.g.\ cf.\ Fig.\ 2 
of Grigahc\`ene et al.\ 2010) between the two types of mode. How can the 
frequency gap be filled?

Two scenarios appear viable. The first has been proposed long ago by 
Balona \& Dziembowski (1999, their Fig.\ 1). Modes with high spherical 
degrees ($l\geq6$) can fill this frequency gap. It has hitherto 
been assumed that such modes cannot be seen in disk-integrated 
photometric measurements, but with the highly improved quality thanks to 
space observations, this assumption must be abandoned. 
Daszy{\'n}ska-Daszkiewicz et al.\ (2002) have shown that, for instance, 
the photometric amplitude of an $l=8$ mode still is 0.3\% of a radial 
mode with the same intrinsic amplitude. Such modes can therefore fill 
the frequency gap.

Another possibility to fill the gap is rapid rotation, illustrated in 
Fig.\ 3. At low rotation speeds, the g and p mode spectra are well 
separated. However, with increasing rotation, the g mode regions extends 
and the p modes become rotationally split predominantly to lower 
frequency. Because of the small vertical amplitude of the g modes, they 
hardly experience the centrifugal force and prograde modes increase in 
frequency with rotation. On the other hand, the mostly vertically moving 
p modes feel the centrifugal force and retrograde modes attain lower and 
lower frequencies. For clarity of presentation, the current example was 
restricted to $l=2$ modes; the effect becomes stronger for higher $l$ 
and with inwardly increasing rotation rate (Dziembowski \& Pamyatnykh 
2008).

\begin{figure}
\plotfiddle{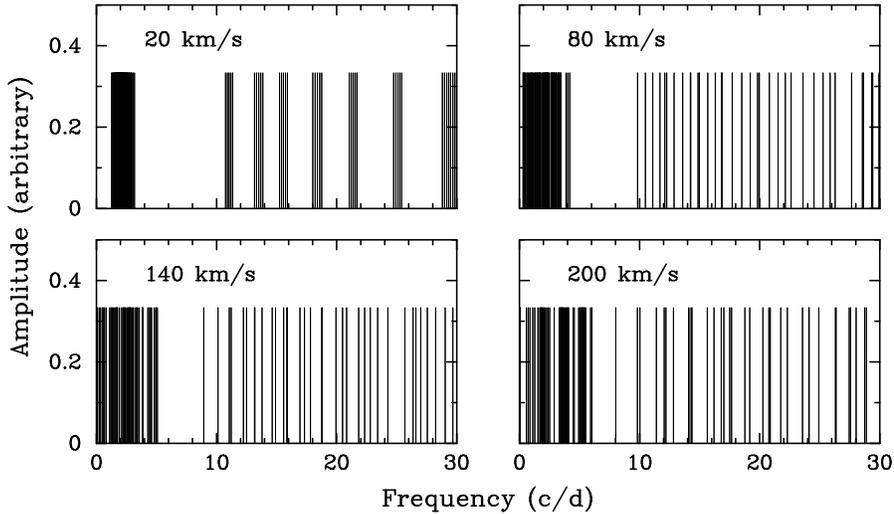}{5.9cm}{00}{82}{82}{-197}{-146}
\caption{Upper four panels: schematic theoretical $l=2$ mode spectrum of 
a Delgam Scudor star rotating with 20 km/s. The g modes are clustered at 
low frequency, the p modes occupy the domain $>10$\,c/d. In consequent 
panels, the model rotates faster and faster. At rotational velocities 
approaching 200 km/s the frequency gap between the two sets of mode 
spectra becomes filled.}
\end{figure}

We are only at the beginning of the exploration of the scientific 
information present in the stellar mode spectra provided from the new 
high-quality measurements. There are several questions that require 
further thought and investigations, such as: what are the decisive 
quantities that separate hybrid pulsators from ``pure'' oscillators of a 
single class? What are the signals in the gap between the g and p/mixed 
mode domains?

Of general importance is the following question: are all the variability 
frequencies we see in these stars due to normal pulsation modes or are 
there other causes? Rotational modulation of light curves can generate 
signals with frequencies in the $\gamma$~Doradus domain. In fact, a 
close connection between rotation and $\gamma$~Doradus pulsation has 
recently been proposed (Balona et al.\ 2011b). Another possibility is 
the presence of r modes in some oscillation spectra (Kurtz et al.\ 
2011).

Given the presence of stellar surface convection in stars within the 
$\delta$~Scuti instability strip, Kallinger \& Matthews (2010) proposed 
that granulation noise may be observed in the amplitude spectra of stars 
with apparently ``too many'' pulsation frequencies. Amplitude/phase 
modulation of pulsation modes may also generate spurious signals in 
frequency analyses. Because of the many possible previously unobserved 
phenomena in stellar light variations, it remains to be discussed 
whether or not there is much point in keeping the designations of 
variable star classes as they are.

\subsection{Chemically peculiar Delgam Scudor stars}

As the first Delgam Scudor stars were discovered, there was the 
hypothesis that their nature could be related to chemical peculiarity, 
as several of these seemed to be Am stars (Matthews 2007). Recent 
spectroscopic analyses (Hareter et al.\ 2011) argue against this 
hypothesis.

The Kepler mission also revealed some Delgam Scudor stars among Ap stars 
and candidates. The detections of the oscillations are quite convincing, 
what remains to be proven is whether these objects are true Ap stars 
(Balona et al.\ 2011a); spectroscopic observations need to be obtained.

\section{New types of hybrid pulsators}

Recently, a hybrid between a $\gamma$~Doradus and a roAp star 
has been reported and modelled (Balona et al.\ 2011). Unfortunately 
the presumed $\gamma$~Doradus mode spectrum is sparse, wherefore 
utilizing both types of mode is not yet possible.

Kepler observations also supplied the first detection of solar-like 
oscillations in the $\delta$~Scuti star HD 187547 (Antoci et al.\ 2011). 
The light variations of the star are dominated by low-order p mode 
pulsations, whereas at higher frequencies the comb-like structure of 
high radial overtone oscillations is observed. Figure 4 shows a 
comparison of the temporal behaviour of a $\delta$~Scuti mode, a 
stochastically driven pulsation and a solar mode. The similarity between 
the latter two is striking.

\begin{figure} 
\plotfiddle{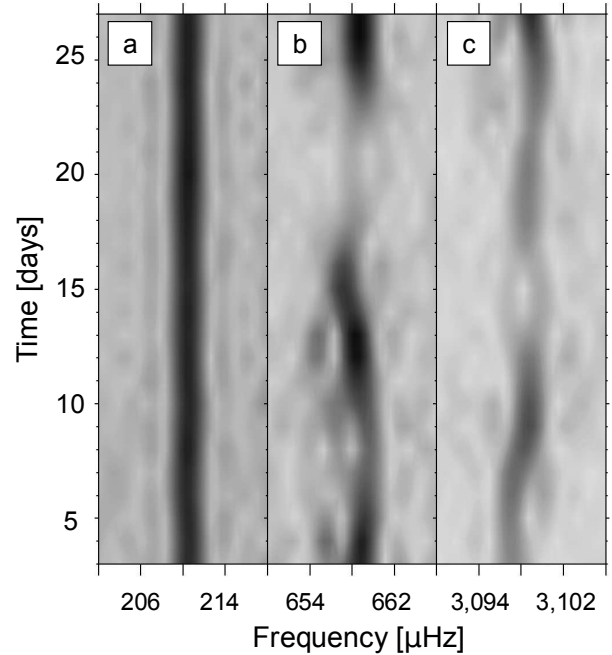}{8.0cm}{00}{90}{90}{-260}{-275} 
\caption{Example Time-Fourier spectra of HD 187547 and the Sun. a)
opacity-driven mode showing temporal stability in the $\delta$ Sct 
frequency region of HD~187547. b) Stochastic mode observed in HD 
187547, displaying erratic behaviour as expected for solar-like 
oscillations. c) A stochastic oscillation mode observed in the Sun with 
the SOHO VIRGO instrument. The data set has the same length and 
sampling as for HD 187547. Reproduced with permission from Antoci et 
al.\ (2011).}
\end{figure}

The importance of the detection of these oscillations lies not only in 
the possibility to sound different interior regions of the star 
asteroseismically. It also implies that stellar surface convection can 
still be efficient in stars about twice as massive as the Sun, where the 
superficial convection zones are as thin as 1\% of the stellar radius. 
The solar-like mode amplitudes and lifetimes can be used to constrain 
convection models in a hitherto observationally unexplored domain of 
physical parameter space, and it will be interesting to see how this 
result can be reconciled with others implying non-efficient surface 
convection in $\delta$ Scuti stars (Lenz et al.\ 2008).

\section{Summary and discussion}

The $\delta$ Scuti stars, the roAp stars, the $\gamma$ Dor stars and 
solar like oscillators all overlap at the intersection of the classical 
instability strip and the main sequence. Each of these classes of 
pulsating star are driven in different regions of the star. To date, we 
know hybrid pulsators between: $\delta$ Scuti and $\gamma$ Dor stars, 
roAp stars and $\gamma$ Dor stars as well as $\delta$ Scuti stars and 
solar-like oscillators. The other three possible combinations of two 
classes of pulsator have not yet been observed.

This status makes sense from the point of view of pulsational driving 
(as we believe to understand it to date), because the mechanisms 
operating in the known hybrids, or their basic physical characteristics, 
do not affect each other negatively. On the other hand, the absence of 
roAp/solar-like hybrids is also not a surprise, as the strong magnetic 
fields of Ap stars are believed to suppress surface convection. Similar 
arguments may be invoked to explain the non-detection of roAp/$\delta$ 
Scuti hybrids: the diffusion and settling of chemical elements is 
supposed to deplete the driving zone for $\delta$ Scuti oscillations of 
the required Helium, and slow wave leakage due to strong magnetic fields 
is believed to provide additional damping.

What about $\gamma$ Doradus/solar-like hybrids? Such stars have not been 
reported to date, but the author believes this is only a question of 
time. Along those lines, given the large fraction of Delgam Scudor 
stars, one may also expect the existence of $\delta$ Scuti/$\gamma$ 
Doradus/solar-like hybrids, which would be another goldmine for 
asteroseismology of main sequence stars.

\acknowledgements I would like to express my deepest respect for the 
people of Japan because of their prudence in dealing with a national 
tragedy. I thank Hiromoto Shibahashi for enouraging everyone to attend 
this meeting despite the difficult circumstances under which it was 
held, and for a generous travel grant. This work was partially supported 
by the Austrian Fonds zur F\"orderung der wissenschaftlichen Forschung 
under grant P20526-N16. Victoria Antoci, Patrick Lenz and Alosha 
Pamyatnykh provided comments that made this article a better one.


\end{document}